\documentclass[aps,pre,twocolumn,floatfix,showpacs,citeautoscript,longbibliography,hyperlinks]{revtex4-1}

\newcommand{\revision}[1]{{{#1}}}

\usepackage{graphicx}
\usepackage{dcolumn}
\usepackage{bm}
\usepackage{bbold}
\usepackage{mathtools}
\usepackage{color}
\usepackage{transparent}
\usepackage[colorlinks=true,citecolor=blue]{hyperref}

\begin{document}
\title{Determination of universal critical exponents using Lee-Yang theory}
\author{Aydin Deger}
\affiliation{Department of Applied Physics, Aalto University, 00076 Aalto, Finland}
\author{Christian Flindt}
\affiliation{Department of Applied Physics, Aalto University, 00076 Aalto, Finland}

\begin{abstract}
Lee-Yang zeros \revision{are points in the complex plane of an external} control parameter at which the partition function vanishes for a many-body system of finite size. In the thermodynamic limit, the Lee-Yang zeros approach the critical value on the real-axis, where a phase transition occurs. Partition function zeros have for many years been considered a purely theoretical concept, however, the situation is changing now as Lee-Yang zeros have been determined in several recent experiments. Motivated by these developments, we here devise a direct pathway from measurements of partition function zeros to the determination of critical points and universal critical exponents of continuous phase transitions. To illustrate the feasibility of our approach, we extract the critical exponents of the Ising model in two and three dimensions from the fluctuations of the total energy and the magnetization in \revision{lattices of finite size}. Importantly, the critical exponents can be determined even if the system is away from the phase transition. \revision{Moreover, in contrast to standard methods based on Binder cumulants, it is not necessary to drive the system across the phase transition.} As such, our method provides an intriguing perspective for investigations of phase transitions that may be hard to reach experimentally, for instance at very low temperatures or at very high pressures.
\end{abstract}

\maketitle

\section{Introduction} Phase transitions are characterized by the abrupt change of a many-body system from one state of matter to another as an external control parameter is varied \cite{Kardar2007,McCoy2009,Domb1984}. In their seminal works, Lee and Yang developed a rigorous theory of phase transitions based on the zeros of the partition function in the complex plane of the control parameter, for instance the fugacity or an external magnetic field \cite{Yang1952a,Lee1952,Blythe2003,Bena2005}. The crucial insight of Lee and Yang was that the partition function zeros with increasing system size will approach the real value of the control parameter for which a phase transition occurs. These ideas are now considered a theoretical cornerstone of statistical physics, and they have found applications across a wide range of topics, including protein folding~\cite{Lee2013,Lee2013a}, percolation \cite{Arndt2001,Dammer2002,Krasnytska2015,Krasnytska2016}, and Bose-Einstein condensation~\cite{Muelken2001,Dijk2015}.

Despite these developments, partition function zeros were for a long time considered a purely theoretical concept. This situation is changing now as Lee-Yang zeros have been determined in several experiments~\cite{Binek1998,Wei2012,Wei2014,Peng2015,Flindt2013,Brandner2017,Deger2018,Flaschner2017}. Recently, partition function zeros were measured using carefully engineered nano-structures involving the precession of interacting molecular spins~\cite{Wei2012,Wei2014,Peng2015}, Cooper pair tunneling in superconducting devices~\cite{Flindt2013,Brandner2017,Deger2018}, or fermionic atoms in driven optical lattices~\cite{Flaschner2017}. In parallel with these experiments, several theoretical proposals have been put forward for the detection of partition function zeros \cite{Gnatenko2017,Wei2017,Gnatenko2018,Kuzmak2019,Krishnan2019}. \revision{These advances motivate further investigations of the information that can be extracted from the determination of Lee-Yang zeros in systems of finite size} and how future experiments on scalable many-body systems may improve our understanding of phase transitions.

\begin{figure}
  \centering
\includegraphics[width=\columnwidth]{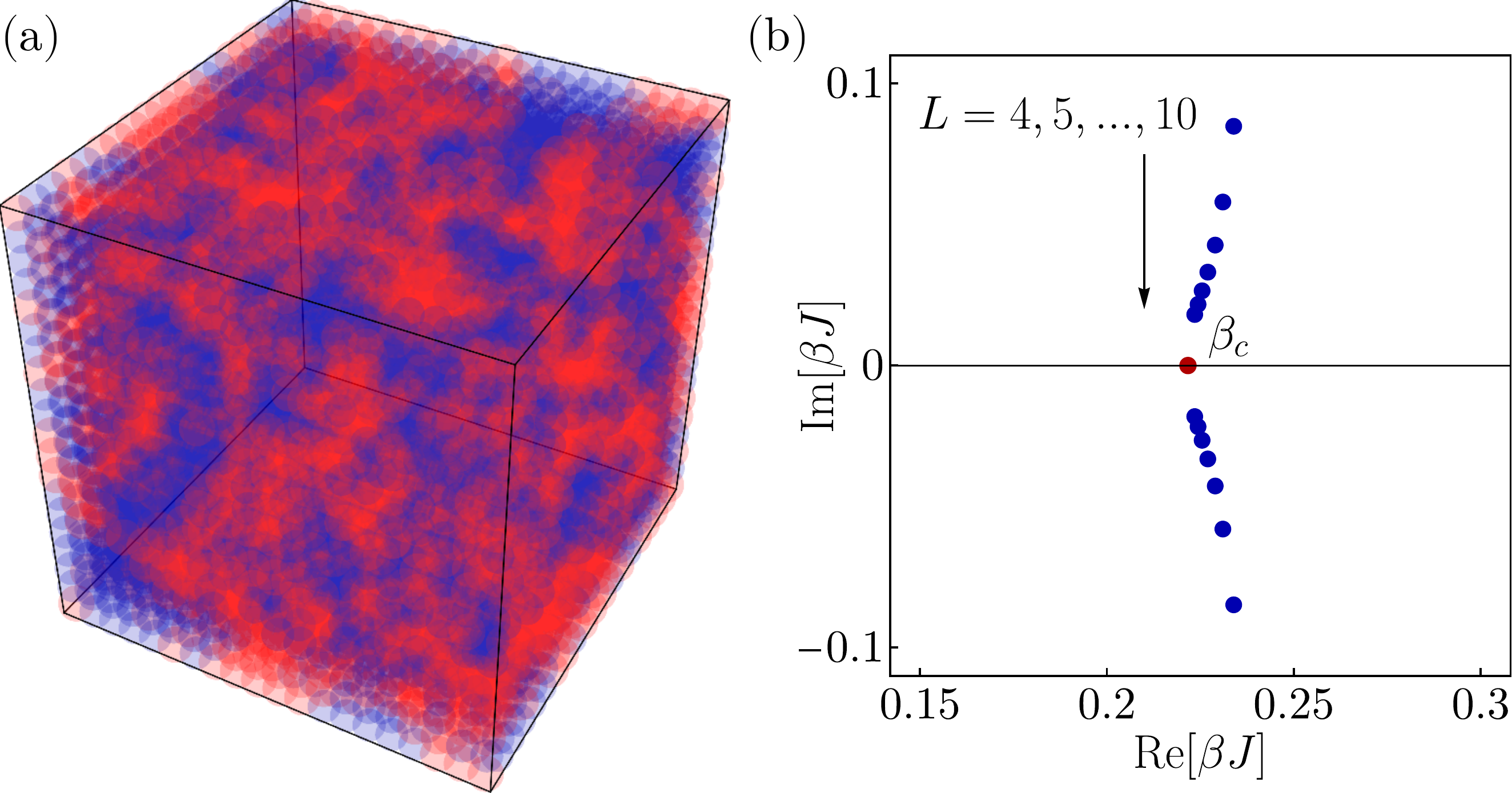}
  \caption{Ising lattice and Fisher zeros. (a) The Ising model, here in $d=3$ dimensions with linear size $L=20$ and $N=L^d=8000$ lattice sites. The color of each site denotes the orientation of its spin, blue (up) or red (down). (b)~From the energy fluctuations, we find the leading partition function zeros in the complex plane of the inverse temperature using Eq.~(\ref{eq:cumuEngMethod}). The inverse temperature is $\beta J=0.23$, where $J$ is the coupling between neighboring spins. No magnetic field is applied. The Fisher zeros approach the critical inverse temperature $\beta_c$ with increasing system size, $L=4,5,\ldots,10$.  Importantly, from the scaling of the Fisher zeros, we can determine the critical exponents as shown in Fig.~\ref{fig:fig2}.}
  \label{fig:fig1}
\end{figure}

In this work, we present a direct pathway from the detection of partition function zeros by measuring or simulating fluctuating observables in systems of finite size to the determination of critical points and universal critical exponents of continuous phase transitions~\cite{Kardar2007,McCoy2009,Domb1984}. Our method combines ideas and concepts from finite-size scaling analysis \cite{Bruce1981,Binder1981,Binder1981a,Binder1997} with the Lee-Yang formalism \cite{Yang1952a,Lee1952,Blythe2003,Bena2005} and theories of high cumulants~\cite{dingle1973,berry2005,Flindt2009,Flindt2010}, and it can be applied in experiments on a variety of phase transitions including non-equilibrium situations such as space-time phase transitions in glass formers~\cite{Garrahan2007,Hedges2009} and dynamical phase transitions in quantum many-body systems after a quench~\cite{Heyl2013,Zvyagin2016,Heyl2017}. Specifically, we determine the partition function zeros from fluctuations of thermodynamic observables and find the critical exponents from the approach of the zeros to the critical value on the real-axis. As a paradigmatic application, we determine the critical points and the universal critical exponents of the Ising model from the fluctuations of energy and magnetization in small lattices. Unlike most conventional methods\revision{, based for instance on Binder cumulants \cite{Binder1981,Binder1981a,Binder1997}}, which require the control parameter to be tuned across the phase transition, we can determine the critical exponents even if the system is away from the phase transition, for example at a \revision{fixed high temperature. (In the Appendices, we discuss the statistical aspects of our method, and we compare it with the use of Binder cumulants}.) As such, our method provides an intriguing perspective for investigations of phase transitions that may be hard to reach experimentally, for instance at very low temperatures or at very high pressures~\cite{Wigner1935,Dias2017}. Moreover, our method opens an avenue for bottom-up experiments on phase transitions, in which nano-scale structures are carefully assembled, for example by adding single spins to an atomic chain on a surface \cite{Choi2019} or by loading individual atoms into an optical lattice one at a time \cite{Eckardt2017}, to increase the system size in a controllable manner.

\section{Ising lattice and criticality}
Figure \ref{fig:fig1}a illustrates the Ising lattice that we consider in this work. The lattice has $N=L^d$ sites, where $L$ is the linear size and $d=2,3$ denotes the spatial dimension. Each site hosts a classical spin which can take on the values $\sigma_i=\pm1$. An external magnetic field of magnitude $h$ can be applied, and neighboring spins are coupled via a ferromagnetic interaction of strength $J>0$. The total energy corresponding to a specific spin configuration $\bm{\sigma}=\{\sigma_i\}$ is then
\begin{equation}
U(\bm{\sigma}) = -J\sum_{\langle i,j\rangle}\sigma_i\sigma_j-h\sum_{i}\sigma_i,
\end{equation}
where the brackets $\langle i,\!j\rangle$ denote summation over nearest-neighbor spins. The thermodynamic properties of the lattice are fully encoded in the partition function
\begin{equation}
\label{eq:partition_function}
Z(\beta,h)=\sum_{\bm{\sigma}} e^{-\beta U(\bm{\sigma})},
\end{equation}
where $\beta=1/(k_B T)$ is the inverse temperature. Phase transitions are signalled by values of the control parameters for which the scaled free energy $f(\beta,h)=-[\ln Z(\beta,h)]/(N\beta)$ becomes non-analytic in the thermodynamic limit of large lattices \cite{Kardar2007,McCoy2009,Domb1984}. The partition function also captures fluctuations of thermodynamic observables. For instance, energy fluctuations can be characterized by the moments $\langle U^n\rangle =[\partial_{-\beta}^nZ(\beta,h)]/Z(\beta,h)$ or cumulants $\langle\!\langle U^n\rangle\!\rangle =\partial_{-\beta}^n \ln Z(\beta,h)$, which follow upon differentiation with respect to the conjugate variable, here the inverse temperature. The moments and cumulants of the magnetization are given in a similar manner by differentiation with respect to the magnetic field strength.

The Ising model exhibits a continuous phase transition, which close to the critical inverse temperature $\beta\simeq \beta_c$ can be completely characterized by a few critical exponents that are independent of microscopic details and are determined solely by general features such as the dimensionality of the problem and its universality class~\cite{Kardar2007,McCoy2009,Domb1984}. As such, the determination of critical exponents is of key importance in statistical mechanics. In the vicinity of the critical point, we may assume that the probability distribution for the total energy obeys the scaling relation $P_L(U)=L^{-1/\nu} f_\infty(U L^{-1/\nu})$, where $f_\infty$ is a scaling function and the critical exponent $\nu$ describes the divergence of the correlation length as we approach the critical temperature \cite{Bruce1981,Binder1981,Binder1981a,Binder1997}. After some algebra, we then obtain scaling relations for the cumulants of the form
\begin{align}
\label{eq:scalingU}
\langle\!\langle U^n \rangle\!\rangle = L^{n/\nu} u_n,
\end{align}
where the $u_n$'s depend only weakly on the system size. As we will see, these relations carry over to the partition function zeros and their approach to the critical point.

\begin{figure*}
  \centering
\includegraphics[width=0.93\textwidth]{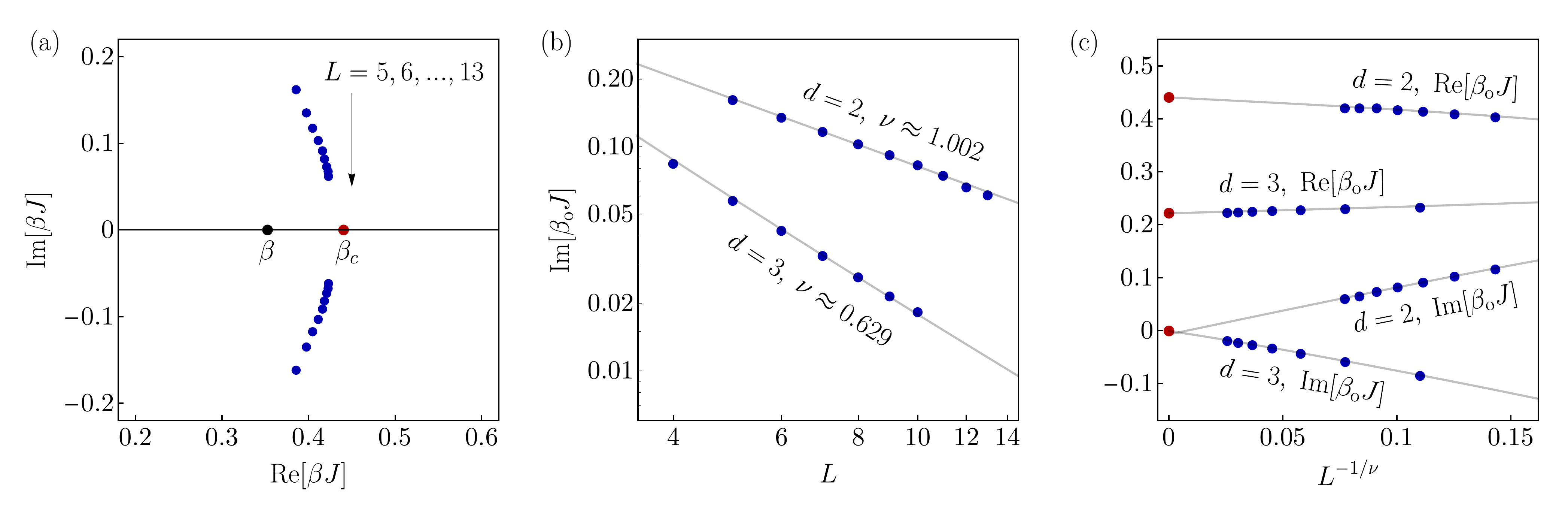}
  \caption{Fisher zeros and critical exponents. (a) The leading Fisher zeros (blue circles) for the Ising model with $d=2$ are extracted from the energy cumulants of order $n=6,7,8,9$. With increasing system size, the Fisher zeros approach the critical inverse temperature $\beta_cJ\simeq 0.4404$ (red circle), which is close to the exact result $\beta_{2\mathrm{D}}J=\ln(1+\sqrt{2})/2\simeq 0.4407$. The simulations were carried out at a temperature above the phase transition, $\beta J=0.35$ (black circle). For the Ising model in Fig.~\ref{fig:fig1} with $d=3$, we find $\beta_cJ= 0.22169$, which is close to the best numerical estimate of $\beta_{3\mathrm{D}}J\simeq 0.22165$. The critical inverse temperatures are determined in panel c. (b) The extracted critical exponents $\nu$ from the finite-size scaling of the imaginary parts are close to the known values for the Ising model, $\nu_{2\mathrm{D}}=1$ (exact) and $\nu_{3\mathrm{D}}\simeq 0.630$ (numerics) \cite{abc}. (c) In the thermodynamic limit, the imaginary part of the zeros vanishes, and the real parts approach the critical values indicated with red circles in panels a and~c.}
  \label{fig:fig2}
\end{figure*}

\section{Partition function zeros and finite-size scaling}
Following the seminal ideas of Lee and Yang, we consider the zeros of the partition function in the complex plane of the control parameter \cite{Yang1952a,Lee1952,Blythe2003,Bena2005}. For finite-size lattices, the partition function is analytic and it can be factorized as
\begin{equation}
\label{eq:parZeros}
Z(\beta,h)=Z(0,h)e^{\beta c}\prod_k\left(1-\beta/\beta_k\right),
\end{equation}
where $\beta_k$ are the zeros in the complex plane of the inverse temperature and $c$ is a constant. The zeros come in complex conjugate pairs, since the partition function is real for real values of $\beta$. Often these zeros are referred to as Fisher zeros, while zeros for complex external fields are known as Lee-Yang zeros. With increasing system size, the partition function zeros approach the real value of the control parameter for which a phase transition occurs in the thermodynamic limit. From the definition of the cumulants, we now obtain the relation \cite{Flindt2013,Brandner2017,Deger2018}
\begin{equation}  \label{eq:cumulantsU}
\langle\!\langle U^n \rangle\!\rangle=(-1)^{(n-1)} \sum_k\frac{(n-1)!}{(\beta_k-\beta)^n}, \ n>1,
\end{equation}
between the cumulants and the partition function zeros. We then see that the high cumulants are mainly determined by the pair of Fisher zeros, $\beta_{\rm o}$ and $\beta_{\rm o}^*$, that are closest to the actual inverse temperature $\beta$ on the real-axis. The contributions from other zeros are suppressed with the distance to $\beta$ and the cumulant order~$n$ \cite{dingle1973,berry2005,Flindt2009,Flindt2010}. Moreover, close to criticality, we expect the scaling relations (\ref{eq:scalingU}) to hold and thus that the leading zeros must approach the critical inverse temperature as~\cite{Itzykson1983,Janke2001,Janke2002}
\begin{equation}
|\beta_{\rm o}-\beta_c| \propto L^{-1/\nu}
\end{equation}
and
\begin{equation}
\label{eq:finitesizeIm}
\mathrm{Im}[\beta_{\rm o}]\propto L^{-1/\nu},
\end{equation}
since the critical inverse temperature is real. These relations are important as they allow us to obtain the critical exponent $\nu$ from the partition function zeros.

\section{Fisher zeros and critical exponents}
Partition function zeros have recently been experimentally determined \cite{Binek1998,Wei2012,Wei2014,Peng2015,Flindt2013,Brandner2017,Deger2018,Flaschner2017}. Lee-Yang zeros have been determined by measuring the quantum coherence of a probe spin coupled to an Ising-type spin bath \cite{Wei2012,Wei2014,Peng2015}, and Fisher zeros have been extracted for a dynamical phase transition involving fermionic atoms in a driven optical lattice \cite{Flaschner2017}. Partition function zeros have also been obtained from the fluctuations of the number of transferred particles in an experiment on full counting statistics of Cooper pair tunneling \cite{Flindt2013,Brandner2017,Deger2018}. Here, we first determine the Fisher zeros of the Ising lattice from fluctuations of the energy, since the energy is conjugate to the inverse temperature. To this end, Eq.~(\ref{eq:cumulantsU}) can be solved for high orders, $n\gg1$, to yield the expression
\begin{equation}
\label{eq:cumuEngMethod}
\begin{bmatrix}
-2\mathrm{Re}[\beta_{\rm o}-\beta]\\
|\beta_{\rm o}-\beta|^2
\end{bmatrix}=\begin{bmatrix}
1& -\frac{\mathsf{\kappa}_{n}^{(+)}}{n}\\
1& -\frac{\mathsf{\kappa}_{n+1}^{(+)}}{n+1}
\end{bmatrix}^{-1}
\begin{bmatrix}
(n-1) \mathsf{\kappa}_{n}^{(-)}\\
n \ \mathsf{\kappa}_{n+1}^{(-)}
\end{bmatrix}
\end{equation}
for the position of the leading partition function zeros, $\beta_{\rm o}$ and $\beta_{\rm o}^*$, in terms of the ratios $\mathsf{\kappa}_{n}^{(\pm)} \equiv \langle\!\langle U^{n\pm 1}\rangle\!\rangle / \langle\!\langle U^{n}\rangle\!\rangle$ of cumulants of subsequent orders. \revision{We stress that the energy fluctuations can be measured (or simulated) at a single fixed temperature, and the Fisher zeros can then be determined using Eq.~(\ref{eq:cumuEngMethod}).}

\begin{figure*}
  \centering
\includegraphics[width=0.93\linewidth]{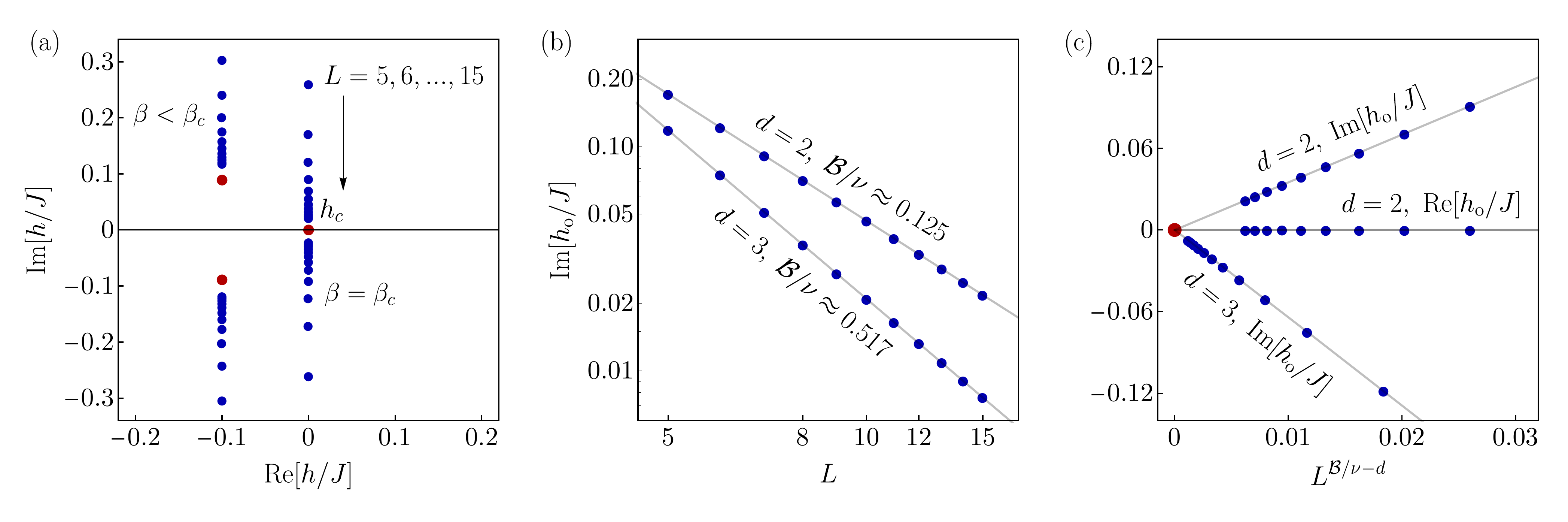}
  \caption{Lee-Yang zeros and critical exponents. (a) The leading Lee-Yang zeros (blue circles) for the Ising model with $d = 2$ are extracted from the magnetization cumulants of order $n = 6, 7, 8, 9$. Above the critical temperature, $\beta=0.8 \beta_c$, the Lee-Yang zeros remain complex in the thermodynamic limit (pair of red circles). For the sake of clarity, these results have been shifted horizontally away from the line $\mathrm{Re}[h/J]=0$. At the critical inverse temperature, $\beta=\beta_c$, the Lee-Yang zeros approach the critical field $h_c=0$ (red circle) with increasing system size. We note that the perpendicular approach to the real-axis shows that the system exhibits a first-order phase transition as a function of the magnetic field \cite{Blythe2003,Bena2005,Biskup2000,Biskup2004}. (b) Finite-size scaling of the imaginary parts of the Lee-Yang zeros and extraction of the ratio of critical exponents $\mathcal{B}/\nu$ for $d=2,3$. (c) Determination of the convergence points of the Lee-Yang zeros (red circle) for $d=2,3$. For $d=3$, the real-part also vanishes (not shown).}
  \label{fig:fig3}
\end{figure*}

To mimic an experiment, we perform Monte-Carlo simulations based on the standard Metropolis algorithm~\cite{Note1, Newman1999}. We thereby evaluate the high cumulants of the energy and subsequently obtain the leading Fisher zeros from Eq.~(\ref{eq:cumuEngMethod}) with increasing system size. The results of this procedure are shown in Fig.~\ref{fig:fig2}a and Fig.~\ref{fig:fig1}b for the Ising lattice in two and three dimensions. Already for small lattices of linear size $L\lesssim 10$, we clearly see that the Fisher zeros approach the critical inverse temperature on the real-axis. A quantitative analysis is provided in Fig.~\ref{fig:fig2}b, where we investigate the finite-size scaling of the imaginary part and extract the critical exponent $\nu$ based on Eq.~(\ref{eq:finitesizeIm}) \footnote{Here we include corrections of the form $\mathrm{Im}[\beta_{\rm o}]\propto L^{-1/\nu}(1+L^{-w})$ as in standard scaling analysis \cite{Itzykson1983}.}. Remarkably, the extracted critical exponents are close to the best-known values for the Ising model in two and three dimensions~\cite{abc}, even if obtained for very small lattices. Moreover, in contrast to conventional methods~\cite{Binder1981,Binder1981a,Binder1997}, which typically require that the control parameter be tuned across the phase transition, we are here able to determine the critical exponents from the energy fluctuations at a fixed temperature above the phase transition. Having determined the critical exponents, we can also find the critical inverse temperature by extrapolating the position of the leading Fisher zeros to the thermodynamic limit in Fig.~\ref{fig:fig2}c. The imaginary part of the Fisher zeros vanishes in the thermodynamic limit, while the real part comes close to the best-known values for the Ising model.

\section{Lee-Yang zeros and critical exponents}
Our method can be applied to a variety of phase transitions, not only in equilibrium settings, but also in non-equilibrium situations such as space-time phase transitions in glass formers~\cite{Garrahan2007,Hedges2009} and dynamical phase transitions in many-body systems after a quench~\cite{Heyl2013,Zvyagin2016,Heyl2017}. (In these cases, the role of the partition function is played by a moment generating function or a return amplitude, which deliver the moments of the fluctuating observable upon differentiation with respect to the appropriate conjugate field.) For example, for the Ising model we may also consider the partition function zeros in the complex plane of the magnetic field. These Lee-Yang zeros can be obtained from the fluctuations of the magnetization similar to how the Fisher zeros are determined using  Eq.~(\ref{eq:cumuEngMethod}). At the critical temperature, the magnetization is assumed to obey the scaling relation $P_L(M)=L^{\frac{\mathcal{B}}{\nu}-d} g_\infty(M L^{\frac{\mathcal{B}}{\nu}-d})$, where $g_\infty$ is a scaling function for the total magnetization and the critical exponent $\mathcal{B}$ describes how the average magnetization vanishes as the critical temperature is approached from below~\cite{Lamacraft2008,Karzig2010,Yin2017,Zhai2018}. This scaling hypothesis translates into scaling relations for the Lee-Yang zeros of the form
\begin{equation}
|h_{\rm o}-h_c| \propto L^{\frac{\mathcal{B}}{\nu}-d}
\end{equation}
and
\begin{equation}
|\mathrm{Im}(h_{\rm o})|\propto L^{\frac{\mathcal{B}}{\nu}-d},
\label{eq:ImScaling}
\end{equation}
where $h_c$ is the magnetic field strength at which the phase transition occurs. We can now determine the Lee-Yang zeros from the simulated fluctuations of the magnetization. The results of this procedure for the Ising lattice with $d=2$ are shown in Fig.~\ref{fig:fig3}a. Above the critical temperature, the Lee-Yang zeros remain complex in the thermodynamic limit, since there is no phase transition. By contrast, at the critical temperature (and also below; not shown), the Lee-Yang zeros reach the real-axis, and we can proceed with the finite-size scaling analysis in Fig.~\ref{fig:fig3}b for $d=2$ and $d=3$. Using Eq.~(\ref{eq:ImScaling}), we then extract the ratio $\mathcal{B}/\nu$ of the critical exponents, also known as the scaling dimension, and again find good agreement with existing estimates. We note that from two independent critical exponents we can obtain all other exponents using the hyperscaling relations derived in renormalization group theory \cite{Pelissetto2002}. Finally, in Fig.~\ref{fig:fig3}c, we show how both the real and imaginary parts of the leading Lee-Yang zeros vanish in the thermodynamic limit, signaling that a phase transition occurs at zero magnetic field.

\begin{figure*}
  \centering
        \includegraphics[width=0.90\textwidth]{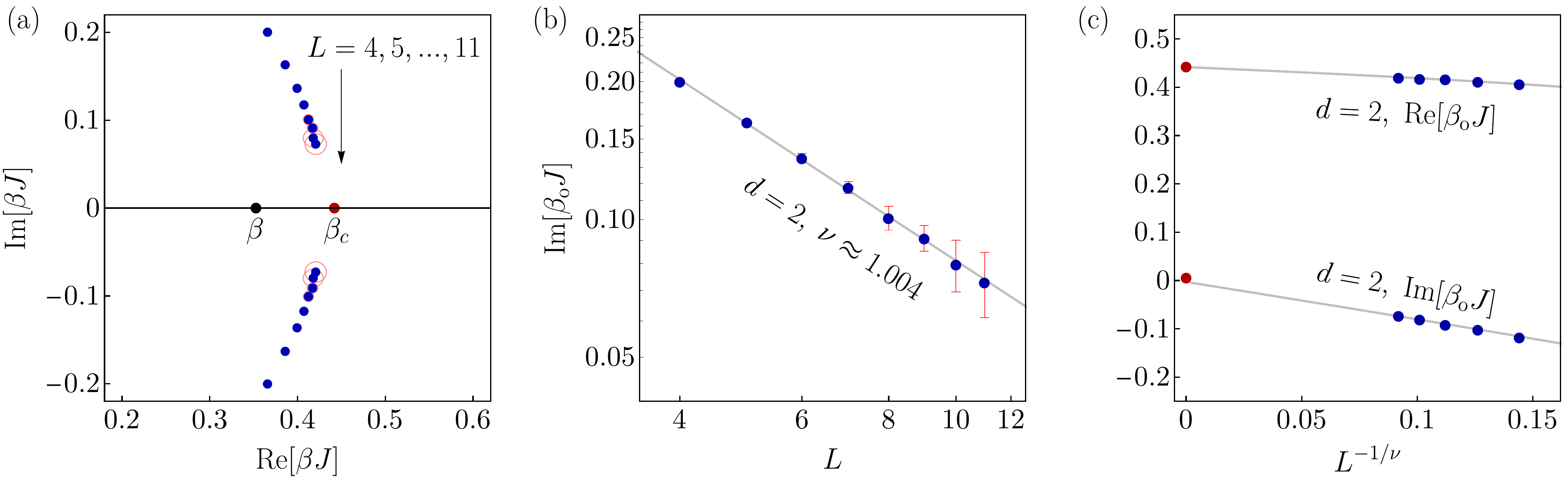}
          \caption{Monte-Carlo simulations and error estimates. (a) The mean location of the leading partition function zeros obtained from $m=10$ Monte Carlo simulations with $10^5$ measurements per site are shown for increasing system sizes, $L=4,5,...,11$. The standard errors are denoted by red ellipses. The partition function zeros (blue circles) are extracted from the energy fluctuations of order $n=5,6,7,8$ at the inverse temperature $\beta=0.8\beta_c$ which is indicated by a black circle. (b) The imaginary parts of the partition function zeros are depicted in a log-log plot as a function of lattice size $L$ together with the errors bars that increase with the system size. The slope of the plot delivers the critical exponent $\nu\approx 1.004$.  (c) Determination of the convergence points of the partition function zeros (red circle). The real part of the zeros moves towards the critical temperature $\beta_c J \approx 0.4419$, which is close to the exact value for the Ising model $\beta_{c}J\approx 0.4407$, whereas the imaginary parts vanish in the thermodynamic limit, signaling a sharp phase transition.
}
  \label{fig:Afig1}
\end{figure*}

\section{Conclusions}
\revision{We have presented a method that makes it possible to identify critical points and determine critical exponents by measuring fluctuations of thermodynamic observables in finize-size systems kept at fixed external control parameters. Our method can not only be applied to equilibrium situations but also non-equilibrium phase transitions, including space-time phase transitions in glass formers and dynamical phase transitions in many-body systems after a quench. We have illustrated the feasibility of our approach using the Ising model for which the critical behavior depends on the dimensionality of the problem as confirmed by our results. Importantly, we can determine the critical points and critical exponents without having to drive the system across the phase transition, which is typically required by other methods. As such, our method paves the way for investigations of phase transitions that may be hard to reach experimentally, for instance at low temperatures. Extending these ideas to the quantum realm constitutes an exciting theoretical challenge for future work.}

\section{Acknowledgements}
We thank K.~Brandner, \revision{F.~Essler}, J.~P.~Garrahan, and \revision{A.~Lamacraft} for insightful discussions. We acknowledge the computational resources provided by the Aalto Science-IT project. Both authors are associated with Centre for Quantum Engineering at Aalto University. The work was supported by the Academy of Finland (projects No.~308515 and 312299). During the final preparations of our manuscript, we became aware of a recent preprint that also investigates partition function zeros for continuous phase transitions \cite{Majumdar2019}.

\begin{appendix}

\section{Monte-Carlo simulations and error estimates}

Here we further discuss the use of our method on the Ising model in $d = 2$ dimensions. As shown in the main text, we can identify the critical points and determine the universal critical exponents by analyzing fluctuating observables for different lattice sizes $N=L^d$ at a single fixed temperature above (or below) the critical point. As such, our method can be applied to a variety of  phase transitions in finite-size systems that are away from the critical temperature. To further analyze the feasibility of our approach, we here discuss the statistical aspects of our method including an error analysis. To this end, we have collected statistics from $m=10$ Monte-Carlo simulations with $10^5$ measurements on an $N$-site Ising lattice using a total of $10^5 \times N$ Monte Carlo steps each. This is two orders of magnitude smaller than for the results presented in the main text. We then determine the leading partition function zeros in the complex plane of the inverse temperature from the cumulants of the energy. The standard errors are calculated as a measure of the precision of the sample means and expressed as $\sigma/\sqrt{m}$, where $\sigma$ denotes the standard deviation. The mean values of $10$ measurements of the partition function zeros are indicated with blue circles and the standard errors are shown by red ellipses in Fig. \ref{fig:Afig1}(a) and vertical error bars in Fig. \ref{fig:Afig1}(b). We stress that our method only requires measurements of the energy at a single fixed temperature, which is indicated by a black circle at $\beta=0.8\beta_{c}$ in Fig.~\ref{fig:Afig1}(a).

\begin{figure*}
  \centering
        \includegraphics[width=0.85\textwidth]{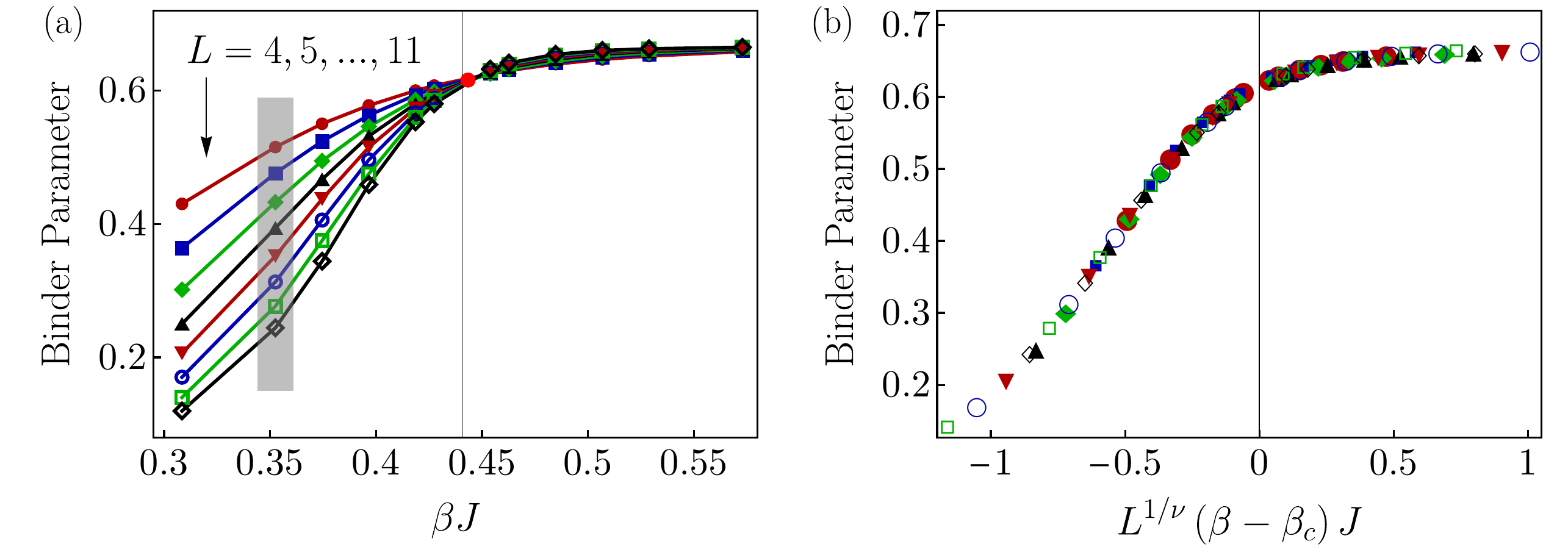}
  \caption{Binder cumulants. (a) The Binder parameter as a function of the inverse temperature for different system sizes. The Binder parameter is evaluated at $\beta=(0.7, 0.8, 0.85, 0.9, 0.95, 0.97, 1.03, 1.05, 1.1, 1.15, 1.2, 1.3)\times \beta_{c}$. The intersection point (red circle) that is extrapolated to the thermodynamic limit, yields the critical temperature $\beta_{c} J \approx0.4434$. The estimation of the critical temperature highly depends upon the proximity of the measurements to the critical point, and collecting more statistics in the vicinity of $\beta_c$ improves the accuracy. The gray frame corresponds to the data that we utilize to determine the critical exponents and critical temperature with our method. (b) Collapse of the Binder cumulant for different system sizes. The estimate of the critical point from (a) is used to tune the critical exponent $\nu\approx 1.07$ so that all data collapse onto a single curve.}
  \label{fig:Afig2}
\end{figure*}

\section{Comparison with Binder cumulants}

We now compare our method to the use of Binder cumulants, which are often used to determine critical points and critical exponents of phase transitions \cite{Binder1981,Binder1981a,Binder1997}. The Binder parameter $M_4$ is a modified fourth-order cumulant of the order parameter, which for Ising models is the magnetization $M$, and it is defined as
\begin{equation}
M_4(L)=1-\frac{\langle M^4 \rangle}{3 \langle M^2 \rangle^2},
\end{equation}
where $\langle M^2 \rangle$ and $\langle M^4 \rangle$ denote the second and fourth order moments of the magnetization. In the thermodynamic limit, the Binder parameter approaches a nontrivial value that depends on the boundary conditions and the lattice structure \cite{Bruce1985, Kamieniarz1993, Selke2005}. The method exploits that $M_4$ depends only weakly on the lattice size exactly at the critical point. Therefore, the crossing point of the Binder parameter for systems of different sizes as a function of the (inverse) temperature yields the critical temperature, where the system undergoes a phase transition in the thermodynamic limit. Based on the scaling behavior of the Binder parameter, one may identify the universality class of the phase transitions by estimating the critical exponent $\nu$ of the correlation length,
\begin{equation}
M_{4}(L)=\widetilde{M}\left(L^{1 / \nu}\left(\beta-\beta_{c}\right)\right),
\end{equation}
where $\widetilde{M}$ is a scaling function. Substituting the critical temperature $\beta_c$, which is obtained from the intersection of the Binder parameters, into this equation causes all curves, corresponding to different system sizes, to collapse onto the same functional form for the correct value of the critical exponent $\nu$. We note that estimating the precise location of the phase transition and the critical exponent requires a detailed study of the finite-size scaling and the use of specialized algorithms, which is beyond the scope of this work. Figure \ref{fig:Afig2} shows the Binder parameter as a function of the temperature and for different system sizes. We note that, in contrast to our method, to make use of Binder cumulants, one must collect statistics of the magnetization at a number of \emph{different} temperatures and tune the system \emph{across} the critical point. From an experimental point of view, it may be a great advantage to work with fixed parameters as we do, for instance if the phase transition takes place at a low temperature, which is hard to reach.
\end{appendix}

\nocite{apsrev41Control}
\bibliographystyle{apsrev4-1}
%
\end{document}